\documentclass[conference]{IEEEtran}
\IEEEoverridecommandlockouts
\usepackage{stmaryrd}
\usepackage{amssymb,amsmath,amsfonts,wasysym,latexsym}
\usepackage[acronyms,shortcuts]{glossaries}
\usepackage[none]{hyphenat}
\usepackage{bm}
\usepackage{tikz}
\usetikzlibrary{patterns,arrows,decorations.pathreplacing}
\usepackage{psfrag}

\usepackage{epstopdf}
\usepackage{amsmath}
\usepackage{scalefnt}
\usepackage[none]{hyphenat}
\usepackage{color}
\usepackage{algorithm}
\usepackage{algpseudocode}
\usepackage{cite}
\usepackage{caption}
\usepackage{subcaption}

\usepackage{caption}
\usepackage{graphicx}
\def\BibTeX{{\rm B\kern-.05em{\sc i\kern-.025em b}\kern-.08em
    T\kern-.1667em\lower.7ex\hbox{E}\kern-.125emX}}

\definecolor{forestgreen}{rgb}{0, 0.5,0.5} 
\definecolor{darkgreen}{rgb}{0,0.392157,0} 
\newcommand{\nmode}[2]{\left[\ma{\mathcal{#1}}\right]_{\left(#2\right)}}
\newcommand{\sumx}[2]{\sum\limits_{#1}^{#2}}
\newcommand{\bb}[1]{\mathbb{#1}}
\newcommand{\ten}[1]{\boldsymbol{\mathcal #1}}
\newcommand{\ma}[1]{\boldsymbol{#1}}
\definecolor{green}{rgb}{0.1,0.75,0.2}

\newacronym{2G}{2G}{second generation}
\newacronym{3G}{3G}{third generation}
\newacronym{4G}{4G}{fourth generation}
\newacronym{5G}{5G}{fifth generation}
\newacronym{B5G}{B5G}{beyond fifth generation}
\newacronym{6G}{6G}{sixth generation}
\newacronym{3GPP}{3GPP}{3$\text{rd}$~Generation Partnership Project}
\newacronym{LTE}{LTE}{long term evolution}
\newacronym{NR}{NR}{new radio}
\newacronym{LS}{LS}{least squares}

\newacronym{IRS}{IRS}{intelligent reconfigurable surface}
\newacronym{RIS}{RIS}{reconfigurable intelligent surface}
\newacronym{LIS}{LIS}{large intelligent surface}
\newacronym{SDS}{SDS}{software-defined surface}

\newacronym{D2D}{D2D}{device-to-device}
\newacronym{BS}{BS}{base station}
\newacronym{UE}{UE}{user equipment}

\newacronym{SU}{SU}{single-user}
\newacronym{MU}{MU}{multi-user}
\newacronym{SISO}{SISO}{single-input single-output}
\newacronym{MISO}{MISO}{multiple-input single-output}
\newacronym{SIMO}{SIMO}{single-input multiple-output}
\newacronym{MIMO}{MIMO}{multiple-input multiple-output}

\newacronym{CSI}{CSI}{channel state information}
\newacronym{LOS}{LOS}{line of sight}
\newacronym{NLOS}{NLOS}{non-line of sight}

\newacronym{QoS}{QoS}{quality-of-service}
\newacronym{SE}{SE}{spectral efficiency}
\newacronym{EE}{EE}{energy efficiency}
\newacronym{SINR}{SINR}{signal to interference plus noise ratio}
\newacronym{SNR}{SNR}{signal to noise ratio}

\newacronym{ProSe}{ProSe}{proximity services}
\newacronym{NSPS}{NSPS}{national security and public safety}

\newacronym{RRM}{RRM}{radio resource management}
\newacronym{MS}{MS}{mode selection}
\newacronym{RA}{RA}{resource allocation}
\newacronym{PC}{PC}{power control}

\newacronym{BCD}{BCD}{block coordinate descent}

\newacronym{RF}{RF}{radio frequency}
\newacronym{AWGN}{AWGN}{additive white Gaussian noise}
\newacronym{MRC}{MRC}{maximum ratio combining}

\newacronym{AF}{AF}{amplify-and-forward}
\newacronym{DF}{DF}{decode-and-forward}

\newacronym{TX}{TX}{transmitter}
\newacronym{RX}{RX}{receiver}
\newacronym{ALS}{ALS}{alternating least squares}
\newacronym{SVD}{SVD}{singular value decomposition}
\newacronym{HOSVD}{HOSVD}{high order singular value decomposition}
\newacronym{THOSVD}{THOSVD}{truncated high order singular value decomposition}
\newacronym{PARAFAC}{PARAFAC}{PARAllel FACtors}
\newacronym{AOD}{AOD}{angle of departure}
\newacronym{AOA}{AOA}{angle of arrival}
\newacronym{URA}{URA}{uniform rectangular array} 
\newacronym{ADR}{ADR}{achievable data rate}
\newacronym{NMSE}{NMSE}{normalized mean square error}
\newacronym{SER}{SER}{symbol error rate}
\newacronym{LRA}{LRA}{low-rank approximation}

\newacronym{ULA}{ULA}{uniform linear array}
\newacronym{mmWave}{mmWave}{milimiter-wave}
\newacronym{CS}{CS}{compressed sensing}
\newacronym{OFDM}{OFDM}{orthogonal frequency division multiplexing}

\begin{document}

\title{IRS Phase-Shift Feedback Overhead-Aware Model Based on Rank-One Tensor Approximation  \\
\thanks{This work was supported by the Ericsson Research, Sweden, and Ericsson Innovation Center, Brazil, under UFC.48 Technical Cooperation Contract Ericsson/UFC. This study was financed in part by the Coordenação de Aperfeiçoamento de
Pessoal de Nível Superior - Brasil (CAPES)-Finance Code
001, and CAPES/PRINT Proc. 88887.311965/2018-00. Andr\'{e}~L.~F.~de~Almeida acknowledges CNPq for its financial support under the grant 312491/2020-4. G. Fodor was partially supported by the Digital Futures project PERCy.}
}

\author{\IEEEauthorblockN{ Bruno Sokal}
\IEEEauthorblockA{\textit{Wireless Research Teleco Group (GTEL)} \\
\textit{Federal University of Ceara (UFC}\\
Fortaleza, Brazil \\
brunosokal@gtel.ufc.br}
\and
\IEEEauthorblockN{ Paulo R. B. Gomes}
\IEEEauthorblockA{\textit{Wireless Research Teleco Group (GTEL)} \\
\textit{Federal University of Ceara (UFC}\\
Fortaleza, Brazil \\
paulo@gtel.ufc.br}
\and
\IEEEauthorblockN{André L. F. de Almeida}
\IEEEauthorblockA{\textit{Wireless Research Teleco Group (GTEL)} \\
\textit{Federal University of Ceara (UFC}\\
Fortaleza, Brazil \\
andre@gtel.ufc.br}
\and
\IEEEauthorblockN{ Behrooz Makki}
\IEEEauthorblockA{\textit{Ericsson Research} \\
\textit{ Ericsson}\\
Göteborg, Sweden \\
behrooz.makki@ericsson.com}
\and
\IEEEauthorblockN{ Gabor Fodor}
\IEEEauthorblockA{\textit{Ericsson Research and KTH Royal Institute of Technology} \\
\textit{Ericsson }\\
Stockholm, Sweden \\
gabor.fodor@ericsson.com}
} 
\maketitle

\begin{abstract}
In this paper, we propose a rank-one tensor modeling \textcolor{black}{approach that yields a compact representation  of the}  optimum \ac{IRS} phase-shift vector \textcolor{black}{for reducing the}  feedback overhead. The main idea consists of factorizing the IRS phase-shift vector as a Kronecker product of smaller vectors, namely factors. {\color{black} The proposed \textcolor{black}{phase-shift} model allows the network to trade-off between achievable data rate and feedback reduction by controling the factorization parameters. }
Our simulations show that the proposed \textcolor{black}{phase-shift} factorization drastically reduces the feedback overhead, while improving the data rate in some scenarios, compared to the state-of-the-art \textcolor{black}{schemes}. 
\end{abstract}

\begin{IEEEkeywords}
IRS, feedback, tensor modeling.
\end{IEEEkeywords}

\section{Introduction}\label{Sec:Introduction}
\Ac{IRS} is a \textcolor{black}{possible candidate} technology for \ac{B5G} and \ac{6G} networks  due to the ability to \textcolor{black}{\textit{control}} the electromagnetic properties of the radio-frequency  propagated waves by performing an intelligent phase-shift to the desired direction \cite{zhang_tutorial,jian2022reconfigurable,rajatheva2020scoring,de2022semi,guo2022dynamic,kenneth_wcnps}. Usually, the \ac{IRS} is defined as a planar ($2$-D) surface with a large number of independent reflective elements in which they can be fully passive or with some elements active \cite{Taha2019,khoshafa2021active,alexandropoulos2020hardware}. The \ac{IRS} is  connected to a smart controller  that sets the desired phase-shift for each reflective element, by applying some bias voltage at the elements e.g., PIN diodes. One main advantage of the fully passive \acp{IRS} is its full-duplex nature, i.e., no noise amplification is observed since no signal processing is possible. However, this fully passive nature \textcolor{black}{makes} the channel state information acquisition process \textcolor{black}{difficult}, since no pilots are processed, thus only the cascade channel can be estimated. Another advantage of an \ac{IRS} with fully passive elements is that the power consumption is concentrated  at the controller. \textcolor{black}{This makes the \ac{IRS}} a more attractive technology in terms of  energy efficient compared to others, \textcolor{black}{e.g.,} amplify-and-forward  and decode-and-forward relays \cite{emil2019_relay}. 

The work of \cite{yang2020intelligent} proposes a protocol design to maximize the transmission rate in IRS-assisted \ac{MIMO}-\ac{OFDM} systems. \textcolor{black}{Also, \cite{Zappone_Overhead_Aware} proposes a framework for a feedback overhead-aware resource allocation in \ac{IRS}-assisted \ac{MIMO} systems. }

In this work, we propose a new overhead-aware model for designing the IRS phase-shifts. \textcolor{black}{Our idea is to represent the IRS phase-shift vector according to a rank-one model. This is achieved by factorizing a \emph{tensorized} version of the IRS phase-shift vector into  a  rank-one tensor, which is modelled as the Kronecker product of a predefined number of factors. } These factors are estimated using  a closed-form solution \textcolor{black}{to} rank-one tensor approximation. After the estimation process, the phases  of the factors are quantized and fed back to the IRS controller, which will reconstruct the IRS phase-shift vector based on \textcolor{black}{its} rank-one  tensor model. The main contributions of the proposed factorization are the following:

 $1)$ \textcolor{black}{The} proposed \textcolor{black}{IRS phase-shift} factorization allows to save network resources. In other words, the network can perform  IRS phase-shift feedback more often, \textcolor{black}{or with a reduced overhead, \textcolor{black}{which  significantly} improves the end-to-end latency}.
 
 $2)$ The proposed method allows a trade-off between the feedback duration and \ac{ADR} performance by \textcolor{black}{properly choosing} the \textcolor{black}{parameters of the rank-one model that represents the IRS phase-shift vector}. \textcolor{black}{This is an important feature of our proposed method, specially for a limited feedback control link.}
 
 $3)$ \textcolor{black}{\textcolor{black}{Our approach} relies on the optimum IRS phase-shift vector, which means that, it can be implemented in \textcolor{black}{different} IRS-assisted systems and in \textcolor{black}{multiple} communication links, i.e.,  downlink or uplink, in \ac{SISO}, \ac{MISO}, \textcolor{black}{as well in} \ac{MIMO} systems.}


\subsection{Notation and Properties} \label{Sec:notation}
Scalars are represented as non-bold lower-case letters $a$, column vectors as lower-case boldface letters $\ma{a}$, matrices as upper-case boldface letters $\ma{A}$, and tensors as calligraphic upper-case letters $\ten{A}$. The superscripts $\{\cdot\}^{\text{T}}$, $\{\cdot\}^{\text{*}}$, $\{\cdot\}^{\text{H}}$ and $\{\cdot\}^{+ }$ stand for transpose, conjugate, conjugate transpose and pseudo-inverse operations, respectively. The operator $\Arrowvert\cdot\Arrowvert_{\text{F}}$ denotes the Frobenius norm of a matrix or tensor, \textcolor{black}{$\bb{E}\{\cdot\}$ is the expectation operator}. The operator $\text{diag}\left(\ma{a}\right)$ converts $\ma{a}$ into a diagonal matrix. \textcolor{black}{Moreover, $\text{vec}\left(\ma{A}\right)$ converts $\ma{A} \in \mathbb{C}^{I_{1} \times R}$ to a column vector $\ma{a} \in \mathbb{C}^{I_{1}R \times 1}$ by stacking its columns on top of each other, while the unvec($\cdot$) operator is the inverse  of the vec operation}. \textcolor{black}{Also,} \textcolor{black}{$\ma{a}_{r} \in \mathbb{C}^{I \times 1}$ represents the $r$-th column of $\ma{A} \in \bb{C}^{I \times R}$}.  The operators $\otimes$ and $\circ$, defines the Kronecker and the outer products, respectively. We make use of the  property 
\begin{align}
\label{eq:p1}\ma{a} \otimes \ma{b} = \text{vec}\left(\ma{b} \circ \ma{a}\right).
\end{align}

%

\section{Tensor Pre-Requisites}\label{Sec:Tensor_background}


Consider a set of matrices $\{\ma{X}_{i_3}\} \in \bb{C}^{I_1 \times I_2}$, for $i_3 = 1,\ldots, I_3$.  By concatenating all $I_3$ matrices to form the third-order tensor $\ten{X} = [\ma{X}_1 \sqcup_3 \ma{X}_2 \sqcup_3 \ldots \sqcup_3 \ma{X}_{I_3} ] \in \bb{C}^{I_1 \times I_2 \times I_3}$, where $\sqcup_3 $ indicates a concatenation in the third dimension. We can interpret $\ma{X}_{i_3}$ as the $i_3$-th frontal slice of $\ten{X}$, defined as $\ten{X}_{..i_3} = \ma{X}_{i_3}$ where the ``$..$" indicates that the dimensions $I_1$ and $I_2$ are fixed. The tensor $\ten{X}$ can be \textcolor{black}{\textit{matricized}} by letting one dimension vary along the rows and the remaining two dimensions along the columns. From $\ten{X}$, we can form three different matrices, referred to as the $n$-mode unfoldings (for $n=\{1,2,3 \}$ in this case), 
\begin{align}
\label{eq:nmode_1}\nmode{X}{1} &= [\ten{X}_{..1},\ldots,\ten{X}_{..I_3}] \in \bb{C}^{I_1 \times I_2I_3}, \\  
\label{eq:nmode_2}\nmode{X}{2} &= [\ten{X}_{..1}^{\text{T}},\ldots,\ten{X}_{..I_3}^{\text{T}}] \in \bb{C}^{I_2 \times I_1I_3}\\ 
\label{eq:nmode_3}\nmode{X}{3} &= [\text{vec}(\ten{X}_{..1}),\ldots,\text{vec}(\ten{X}_{..I_3})]^{\text{T}} \in \bb{C}^{I_3 \times I_1I_2}.
\end{align} 

\subsection{Tensorization}\label{Sec:tensor_background_tensorization}
{\color{black}
The tensorization operation consists  of mapping the elements of a vector into high-order tensors. Let us define the vector $\ma{y} \in \bb{C}^{N \times 1}$, in which $N = \prod\limits_{p=1}^P {N_p}$. By applying the tensorization operator, defined as $\ten{T} \{ \cdot \}$, we can form the tensor $\ten{Y} = \ten{T}\{\ma{y}\}  \in \bb{C}^{N_1 \times N_2 \times \ldots \times N_P}$. The mapping of elements from $\ma{y}$ to $\ten{Y}$ is defined as
\begin{align}
\label{eq:tensozire}
\ten{Y}_{n_1,n_2,\ldots, n_P} = \ma{y}_{n_1 + (n_2 -1)N_1  + \ldots + (n_P - 1)N_{P-1}\cdots N_2N_1},
\end{align}
where $n_p = \{1,\ldots , N_p \}$, for $p = \{1, \ldots ,P\}$. This operator plays a key role on the proposed feedback-aware method where we reshape the elements of the IRS phase-shift vector into a $P$-order tensor.

}
\subsection{Rank-One Tensors}\label{Sec:PARAFAC_decom}

A rank-one matrix can be described as the outer product of two vectors. Similarly to the matrix case, a rank-one tensor is given as the outer product of three or more vectors. For a $P$-order rank-one tensor $\ten{Y} $, it can be written as 
\begin{align}
\label{eq:r1_tensor} \ten{Y} = \ma{a}^{(1)} \circ \ma{a}^{(2)} \circ \ldots \circ \ma{a}^{(P)} \in \bb{C}^{I_1 \times I_2 \times \ldots \times I_P},
\end{align} 
where $\ma{a}^{(p)} \in \bb{C}^{I_p \times 1}$ is the $p$-th factor of $\ten{Y}$, for $p = \{1,\ldots,P\}$. The $p$-th rank-one matrix unfolding of $\ten{Y}$,  defined as $\nmode{Y}{P} \in \bb{C}^{I_p \times I_1 \cdots I_{p-1}I_{p+1} \cdots I_P}$, is given by
\begin{align}
\label{eq:nmode_parafac} \nmode{Y}{p} \hspace*{-0.1cm}=\hspace*{-0.1cm} \ma{a}^{(p)} \hspace*{-0.08cm}\left(\ma{a}^{(P)} \otimes \hspace*{-0.05cm}\ldots \hspace*{-0.05cm}\otimes \ma{a}^{(p+1)} \otimes  \ma{a}^{(p-1)}\ldots \otimes \ma{a}^{(1)} \hspace*{-0.08cm}\right)^{\text{T}} \hspace*{-0.09cm}.
\end{align}

Defining   $ \ma{y} = \text{vec}\left(\ten{Y}\right) \in \bb{C}^{I_1\cdots I_P \times 1}$ and using  (\ref{eq:p1}), we have
\begin{align}
\label{eq:vec_r1_ten} \ma{y} = \ma{a}^{(P)} \otimes \ldots \otimes \ma{a}^{(1)}.
\end{align}

\section{System Model}\label{Sec:System_Model}

We consider an IRS-assisted MIMO system, where the \ac{TX} is equipped with a \ac{ULA} with $M_T$ antenna elements, the \ac{RX} is equipped with \ac{ULA} with $M_R$ antenna elements and the \ac{IRS} has $N$ reflective elements. To simplify \textcolor{black}{the discussions}, let us consider a single stream transmission, assuming that there is no direct link between the \ac{TX} and \ac{RX}. First, the \ac{TX} sends a pilot signal to the \ac{RX} with the aid of the \ac{IRS}. Since there is no signal processing at the \ac{IRS}, the channel estimation and the IRS phase-shifts optimization are performed at the \ac{RX}. The received signal after processing the pilots is given by
\begin{align}
  \label{eq:sig}  y = \ma{w}^{\text{H}}\ma{G}\ma{S}\ma{H}\ma{q} + \ma{w}^{\text{H}}\ma{b},
\end{align}
where $\ma{b} \in \bb{C}^{M_R \times 1}$ is the additive noise at the receiver with  $\mathbb{E}[\ma{b}\ma{b}^{\text{H}}] = \sigma^2_b \ma{I}_{M_r}$, $\ma{w} \in \bb{C}^{M_R \times 1}$ and $\ma{q} \in \bb{C}^{M_T \times 1}$ are the receiver and transmitter combiner and precoder, respectively. $\ma{H} \in \bb{C}^{N \times M_T}$ and $\ma{G} \in \bb{C}^{M_R \times N}$ are the TX-IRS and IRS-RX involved channels, and $\ma{S} = \text{diag}(\ma{s}) \in \bb{C}^{N \times N}$ where $\ma{s} = [e^{j\theta_1}, \ldots, e^{j\theta_N}  ]\in \bb{C}^{N \times 1}$ is the IRS phase-shift vector, and $\theta_n$ is the phase-shift applied to the $n$-th \ac{IRS} element. 

After the channel estimation step, an optimization step of the precoder and combiner (active beamformers) vectors $\ma{w}$ and $\ma{q}$,  and the IRS phase-shift vector $\ma{s}$ (passive beamformer) is performed. Later, the \ac{RX} needs to feedback to the IRS controller the designed phase-shift so that the elements can reconfigure the elements to have the optimal phase-shift. From the fact that this feedback may occur through a limited control channel and that the IRS  may contain several hundreds to thousand of  elements, the feedback of each phase-shift with a certain resolution implies in a signaling overhead.  
 

Related to the issue of the overhead IRS phase-shift signaling feedback, the work of \cite{Zappone_Overhead_Aware} models the feedback duration as

\begin{equation}
    \label{eq:zappone_tf} T_{\text{F}} =  \frac{Nb_\text{F}}{B_{\text{F}} \text{log}\left( 1+ \frac{p_{\text{F}}|{g}_{\text{F}}|^2}{B_\text{F} N_0} \right)},
\end{equation}
where \textcolor{black}{$N$ is the total number of phase-shifts of the IRS to be fed back,} $B_{\text{F}}$, $p_{\text{F}}$ are the feedback bandwith and power, $g_{\text{F}}$ is the scalar control channel used, $b_{\text{F}}$ is the resolution of each phase-shift, and $N_0$ is the noise power density. 

\section{Proposed Feedback-Aware Method}\label{Sec:Proposed_Method}

In this section, we describe our proposed feedback-aware method that focuses on reducing the feedback duration $T_\text{F}$, given in (\ref{eq:zappone_tf}). \textcolor{black}{Suppose} that the channels $\ma{H}$ and $\ma{G}$ are estimated at the \ac{RX}. Then, the $N$ phase-shifts of the IRS are determined based on  state-of-the-art algorithm \cite{Zappone_Overhead_Aware}, and are represented in a vector format as $\ma{s} = [e^{j\theta_1}, e^{j\theta_2}, \ldots, e^{j\theta_N}]  \in \bb{C}^{N \times 1}$. Our idea \textcolor{black}{is to} factorize $\ma{s}$  as the  Kronecker product of $P$ factors, i.e., 
\begin{equation}
    \label{eq:fac_s} \ma{s} =  \ma{s}^{(P)} \otimes \ldots \otimes \ma{s}^{(1)} \in \bb{C}^{N_P \cdots N_1 \times 1}
\end{equation}
where $\ma{s}^{(p)} \in \bb{C}^{N_p \times 1}$ $p$-th  factor. From  (\ref{eq:fac_s}), we have that $N = \prod\limits_{p=1}^P N_p$. The main idea of the proposed model is to feedback to the IRS controller only the phase-shifts of the factors, $\ma{s}^{(p)}$, for $p = \{1,\ldots, P\}$, which results into the feedback of $\sumx{p=1}{P} N_p$ phase-shifts, instead of $N = \prod\limits_{p=1}^P$ phase-shifts \textcolor{black}{in the case with no factorization}.

\begin{figure}[!t]
	\centering\includegraphics[scale=0.08]{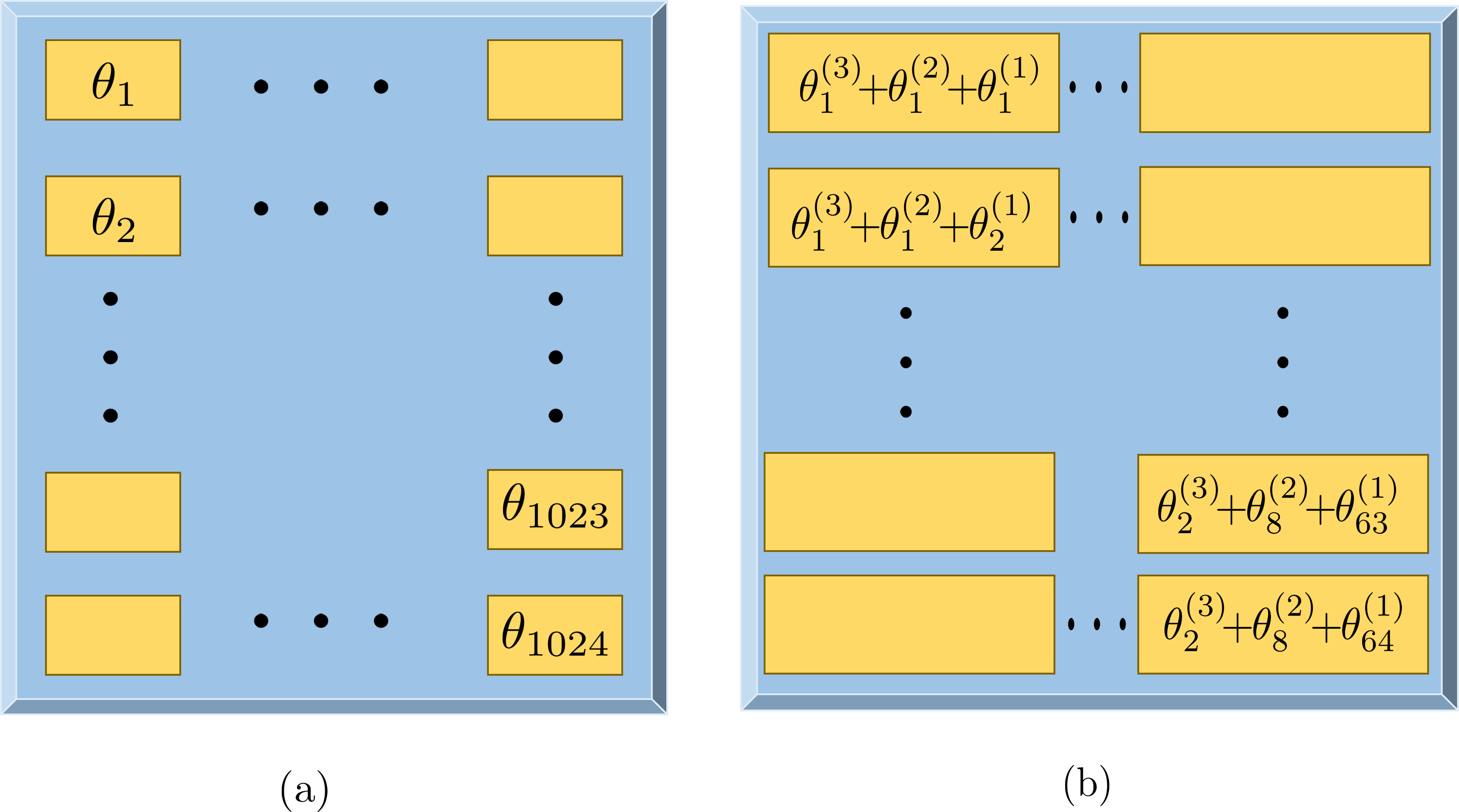}
	\caption{(a) IRS with $N$ phase-shifts, without the factorization, (b) IRS with $N$ phase-shifts factorized into $P=3$ factors. }
	\label{fig:IRS_factorized}
\end{figure}

\textbf{Example:} To illustrate the impact of this factorization on the IRS phase-shift feedback duration, let us consider the simple scenario, illustrated in Figure \ref{fig:IRS_factorized}, where  we have an IRS with $N=1024$ elements. Normally, supposing $N$ independent elements, in Figure \ref{fig:IRS_factorized} (a), $N=1024$ phase-shifts would have to be determined at some system node and conveyed to the IRS controller. On the other hand, considering that, as one example, $P=3$ with the following factors $\ma{s}^{(1)} = [e^{j\theta^{(1)}_1}, \ldots, e^{j\theta^{(1)}_{64}} ] \in \bb{C}^{64 \times 1}$, $\ma{s}^{(2)} = [e^{j\theta^{(2)}_1}, \ldots, e^{j\theta^{(2)}_8} ] \in \bb{C}^{8 \times 1}$ and $\ma{s}^{(3)} = [e^{j\theta^{(3)}_1}, e^{j\theta^{(3)}_2} ] \in \bb{C}^{2 \times 1}$, i.e., $N_1 = 64$, $N_2 =8$ and $N_3=2$, only $N_1 + N_2 + N_3 = 74$ phase-shifts  has to be reported to the IRS controller, drastically reducing the feedback \textcolor{black}{overhead} compared to the traditional scenario (a), \textcolor{black}{by a factor of $\approx 1400 \%$}. $\blacksquare$ 

With the factorization, an additional complexity is introduced on the IRS controller, which will have to build the phase-shift vector, which in this case is given as  $\ma{s} = \ma{s}^{(3)} \otimes \ma{s}^{(2)} \otimes  \ma{s}^{(1)} \in \bb{C}^{1024 \times 1}$. Physically, the Kronecker product in (\ref{eq:fac_s}) represents a summation of the factors phase-shifts, \textcolor{black}{ as shown in Figure \ref{fig:IRS_factorized} (b)}. The proposed feedback duration is given by


\begin{equation}
    \label{eq:feedback_proposed} T_\text{F}^{\text{prop}} = \frac{T_{\text{PR}} +  \sumx{p=1}{P} N_p b^{(p)}_\text{F}}{B_{\text{F}} \text{log}\left( 1+ \frac{p_{\text{F}}|{g}_{\text{F}}|^2}{B_\text{F} N_0} \right)},
\end{equation}
where $T_{PR}$ is the duration of the preamble that informs the values of the factorization parameters, such as the number of factors $P$, the size of each factor $N_p$, and the number of quantization bits used for each factor $b^{(p)}_{\text{F}}$, for $p=\{1,\ldots,P\}$.


The physical implications on the choice of the proposed rank-one tensor approximation parameters, i.e., the number of the factors $P$ and  the size of each factor $N_p$ are discussed in Section \ref{Sec:Parameters}. In a general view, the proposed factorization consists of three steps:

$1)$ \textbf{Rearrangement  of elements}: In this step, the optimum phase-shift vector $\ma{s} \in \bb{C}^{N \times 1}$ is rearranged into a $P$-th order tensor $\ten{S} \in \bb{C}^{N_1 \times N_2 \times \ldots \times N_P}$, with $N = \prod\limits_{p=1}^{P} N_p $.  This is accomplished by mapping the elements of the IRS phase-shift vector $\ma{s}$ into the tensor $\ma{S}$, using the tensorization operator, \textcolor{black}{given in  (\ref{eq:tensozire})}.

$2)$ \textbf{Rank-One tensor approxmation}: In this step, the \ac{RX} estimates the factors of the IRS phase-shift tensor $\ten{S}$. For this, a prior art rank-one tensor estimation algorithmz can be employed, such as the \ac{HOSVD} \cite{Lathauwer2000bestr1r2}, which is described in Algorithm \ref{Alg:HOSVD_IRS}.

$3)$ \textbf{Normalization}: The estimated factors have \textcolor{black}{their} entries normalized to ensure the unitary modulus constraint of the IRS phase-shift vector. This can be achieved by only taking the angles of the estimated factors.

%

\subsection{Rank-One IRS model }\label{Sec:PARAFAC_IRS}
After the tensorization step, the RX will approximate the optimum phase-shift tensor $\ten{S}$ using a rank-one model, i.e.,
\begin{equation}
\label{eq:tenS_parafac}    \ten{S} \approx  \ma{s}^{(1)} \circ \ldots \circ \ma{s}^{(P)} \in \bb{C}^{N_1 \times \cdots \times N_P}.
\end{equation}
Please note that, the approximation comes from the fact that we are fitting independent phase-shifts as a combination of $P$ sets of phase-shifts, thus an approximation error is expected. However, as it will be explained in Section~ \ref{Sec:Simulation_Results}, for scenarios with \textcolor{black}{moderate/strong \ac{LOS} components (approximated rank-one channels), which is the case of interest in IRS networks,} this fitting error is negligible on the \ac{ADR} performance.

The RX estimates the factor components solving the following problem
\begin{align}
  \label{eq:parafac_min} \left[\hspace*{-0.065cm} \ma{\hat{s}}^{(1)}, \ldots ,\ma{\hat{s}}^{(P)}\hspace*{-0.075cm} \right] \hspace*{-0.08cm} =\hspace*{-0.08cm} \underset{\ma{s}^{(1)}, \ldots ,\ma{s}^{(P)}}{\text{argmin}} \left|\left| \ten{S} -  \ma{s}^{(1)} \circ \ldots \circ \ma{s}^{(P)} \right|\right|^2_\text{F},
\end{align}
where $\ma{s}^{(p)} \in \bb{C}^{N_p \times 1}$ is the $p$-th factor. From  (\ref{eq:nmode_parafac}), the $p$-mode unfolding of $\ten{S}$, defined as $\nmode{S}{p} \in \bb{C}^{N_p \times N_1\cdots N_{p-1}N_{p+1} \cdots N_P }$, is the approximated rank-one matrix, given as 
\begin{equation}
 \label{eq:nmode_irs_parafac}   \nmode{S}{p} \approx \ma{s}^{(p)}\left(\ma{s}^{(P)} \otimes \ldots \otimes \ma{s}^{(p+1)} \otimes \ma{s}^{(p-1)} \otimes \ldots \otimes\ma{s}^{(1)} \right)^{\text{T}}.
\end{equation}
 \begin{algorithm}
	\begin{algorithmic}[1]
		\caption{Feedback-Aware Rank-One Approximation}
		\label{Alg:HOSVD_IRS}
		\State \textbf{Inputs}: Tensor $\ten{S}$	
		\For{$p =1:P$}		
			\State Define the SVD of $\nmode{S}{p}$ as $\ma{U}^{(p)}\ma{\Sigma}^{(p)}\ma{V}^{(p)\text{H}}$, compute an estimate \textcolor{black}{of} the $p$-th factor $\ma{s}^{(p)}$ as
			\begin{align*}
\ma{\hat{s}}^{(p)} = e^{j \angle \ma{u}_{.1}^{(p)}}
\end{align*}	
	\EndFor
		\State Return $\ma{\hat{s}}^{(1)}$, $\ldots$, $\ma{\hat{s}}^{(P)}$.
	\end{algorithmic}
\end{algorithm}

To solve the problem in (\ref{eq:parafac_min}),  we make use of the rank-one \ac{HOSVD}, which consists of computing,  a \ac{SVD}\footnote{In fact, since we want the left dominant singular vector of each unfolding, the power method \cite{golub2013matrix} can be used instead of computing the whole \ac{SVD}. } for each unfolding of $\ten{S}$ 
  to extract the left dominant singular vector, i.e., considering the SVD of the $p$-th unfolding of $\ten{S}$, $\nmode{S}{p}$ as $\ma{U}^{(p)}\ma{\Sigma}^{(p)}\ma{V}^{(p)\text{H}}$, the algorithm set an estimation for the $p$-th factor $\ma{s}^{(p)}$ as $\ma{\hat{s}}^{(p)} = e^{j \angle \ma{u}_{.1}^{(p)}}\in \bb{C}^{N_p \times 1}$, where $\ma{u}_{.1}^{(p)}$ is the dominant left singular vector of $\ma{s}^{(p)}$.
%

\subsection{Phase-shift Quantization and Feedback}

It is important to mention that, after the  rank-one factor estimation procedure in Algorithm~\ref{Alg:HOSVD_IRS}, the \ac{RX} quantizes the phase-shifts of each factor  with $b_F^{(p)}$ bits. Let us define $\ma{\tilde{s}}^{(p)} \in \bb{C}^{N_p \times 1}$ as the $p$-factor after the quantization process, i.e., $\ma{\tilde{s}}^{(p)} = \mathcal{Q}\left\{\ma{\hat{s}}^{(p)},b_\text{F}^{(p)}\right\}$.  Then, the phase-shifts of the $P$ factors are conveyed to the IRS controller via a control link channel.  Finally, the IRS controller reconstructs the IRS phase-shift vector as
\begin{align}
\label{eq:IRS_recon}\ma{s} = \ma{\tilde{s}}^{(P)} \otimes \ldots \otimes \ma{\tilde{s}}^{(1)} \in \bb{C}^{N \times 1}.  
\end{align} 

As observed, the proposed method allows the system to adapt the phase-shift resolution of the factors depending on the control link availability. This feature is very important in order to minimize the performance loss, in the case of limited control link.
\subsection{On the Effect of the Factorization Parameters}\label{Sec:Parameters}

In this section, we discuss the choice of the factorization parameters and the system performance implications.

 \textbf{Number of factors} $P$: This parameter defines the total number of factors used in the rank-one tensor approximation.  Its minimum value for the proposed factorization is $P=2$, i.e., the value $P=1$ means that no factorization is employed. By increasing the value of $P$, \textcolor{black}{the number of factors of the factorization model is increased}, allowing to reduce the size of the factor components $N_p$, consequentially, increasing $P$ reduces the phase-shift feedback overhead. 


\textbf{Size of factor components} $N_p$: The size of the factor components indicates the total number of independent phase-shifts in the proposed solution, which it also affects the performance.  For example, for $N=256$ and choosing $P=2$, two possible configurations are $N_1=128$, $N_2=2$ and $N_1=N_2=16$. For the first choice, the system has more independent phase-shifts ($130$), thus a higher spectral efficiency. However, its feedback overhead is higher than that of second configuration that requires only $32$ phase-shifts to be reported in the feedback channel. 
\section{Simulation Results}\label{Sec:Simulation_Results}

In this section, we evaluate the performance of the proposed IRS phase-shift overhead-aware feedback model in terms of feedback duration and \ac{ADR}. For a fair comparison between the proposed rank-one model and the state-of-the-art, we optimize the precoder ($\ma{q}$), combiner ($\ma{w}$), and the IRS phase-shifts ($\ma{s}$) using the upper-bound algorithm of the state-of-the-art \cite{Zappone_Overhead_Aware}. In this case, they are given as
\begin{align*}
\ma{w} = \ma{U}^{(\text{G})}_{.1}, \hspace{0.1cm} 
\ma{q} =  \ma{V}^{(\text{H})}_{.1}, \hspace{0.1cm}
\ma{s}_n = e^{-\angle \left(\ma{V}^{ \text{G})}_{n,1} \cdot \ma{U}^{(\text{H})}_{n,1} \right)}, n = \{1, \ldots, N\},
\end{align*} 
where $\ma{U}^{(\text{G})}_{.1} \in \bb{C}^{M_R \times 1}$, $\ma{V}^{ \text{G})}_{.1} \in \bb{C}^{N \times 1}$ are the dominant left and right singular vectors of $\ma{G}$, while $\ma{U}^{(\text{H})}_{.1} \in \bb{C}^{N \times 1}$, $\ma{V}^{ \text{H})}_{.1} \in \bb{C}^{M_T \times 1}$ are the dominant left and right singular vectors of $\ma{H}$. The \ac{ADR} is given by

\begin{align}
\label{eq:ADR} \text{ADR} = \text{log}_2 \left( 1 + \frac{ \left|\ma{w}^{\text{H}} \ma{G}\ma{S}\ma{H}\ma{q} \right|^2   }{ \sigma^2_b} \right),
\end{align}
where, the noise variance $\sigma^2_b = 0.1$,  $\ma{S} = \text{diag}(\ma{s}) \in \bb{C}^{N \times 1}$ is the matrix that contains in its diagonal the optimized feedback phase-shifts, which, in the proposed approach is given in  (\ref{eq:IRS_recon}).  Regarding the involved channels, in (\ref{eq:sig}), they  are modelled as 

\begin{align}
\label{eq:channel_H} \ma{H} &= \sqrt{\alpha_H \frac{K_H}{K_H+1}} \ma{H}_{\text{LOS}}  +  \sqrt{\frac{1}{K_H+1}} \ma{H}_{\text{NLOS}}, \\
\label{eq:channel_G}\ma{G} &= \sqrt{\alpha_G \frac{K_G}{K_G+1}} \ma{G}_{\text{LOS}}  +  \sqrt{\frac{1}{K_G+1}} \ma{G}_{\text{NLOS}}, 
\end{align}
where $\alpha_H$ and $\alpha_G$ are the path-loss components of the \ac{TX}-IRS and IRS-\ac{RX} links, respectively. $K_H$ and $K_G$ are the Rician factors for the channels $\ma{H}$ and $\ma{G}$, respectively. $\ma{H}_{\text{LOS}}$, $\ma{G}_{\text{LOS}}$ are modelled as geometric-based channels, while the entries of $\ma{H}_{\text{NLOS}}$, $\ma{G}_{\text{NLOS}}$ are modelled as circularly symmetric complex Gaussian random variables, with zero mean an unit variance, i.e., $\ma{H}_{\text{NLOS}} \sim \mathcal{CN}(0,\ma{I}_{M_T})$ and $\ma{G}_{\text{NLOS}} \sim \mathcal{CN}(0,\ma{I}_{M_R})$. \textcolor{black}{ We consider $\alpha_H=\alpha_G = 1$ for experimental purposes, to observe the performance impact of the proposed IRS phase-shift factorization. Also, we consider $K_H=K_G=K$ as metric for performance evaluation.
The details of the channel model are given in Appendix \ref{App:channel_model}.}

Let us define the vector $\ma{N}_{\text{P}} = \left[N_1, \ldots, N_P\right]^{\text{T}} \in \bb{R}^{P \times 1}$ that contains the size of all individual factors, for a certain $P$. Likewise, let us define the vector $\ma{b}^{(\text{P})}_\text{F} = \left[b^{(1)}_\text{F}, \ldots, b^{(P)}_\text{F} \right]^{\text{T}} \in \bb{R}^{P \times 1}$ that contains the number of bits used for quantization in each factor, for a certain $P$.

\begin{figure}[!t]
	\centering\includegraphics[scale=0.55]{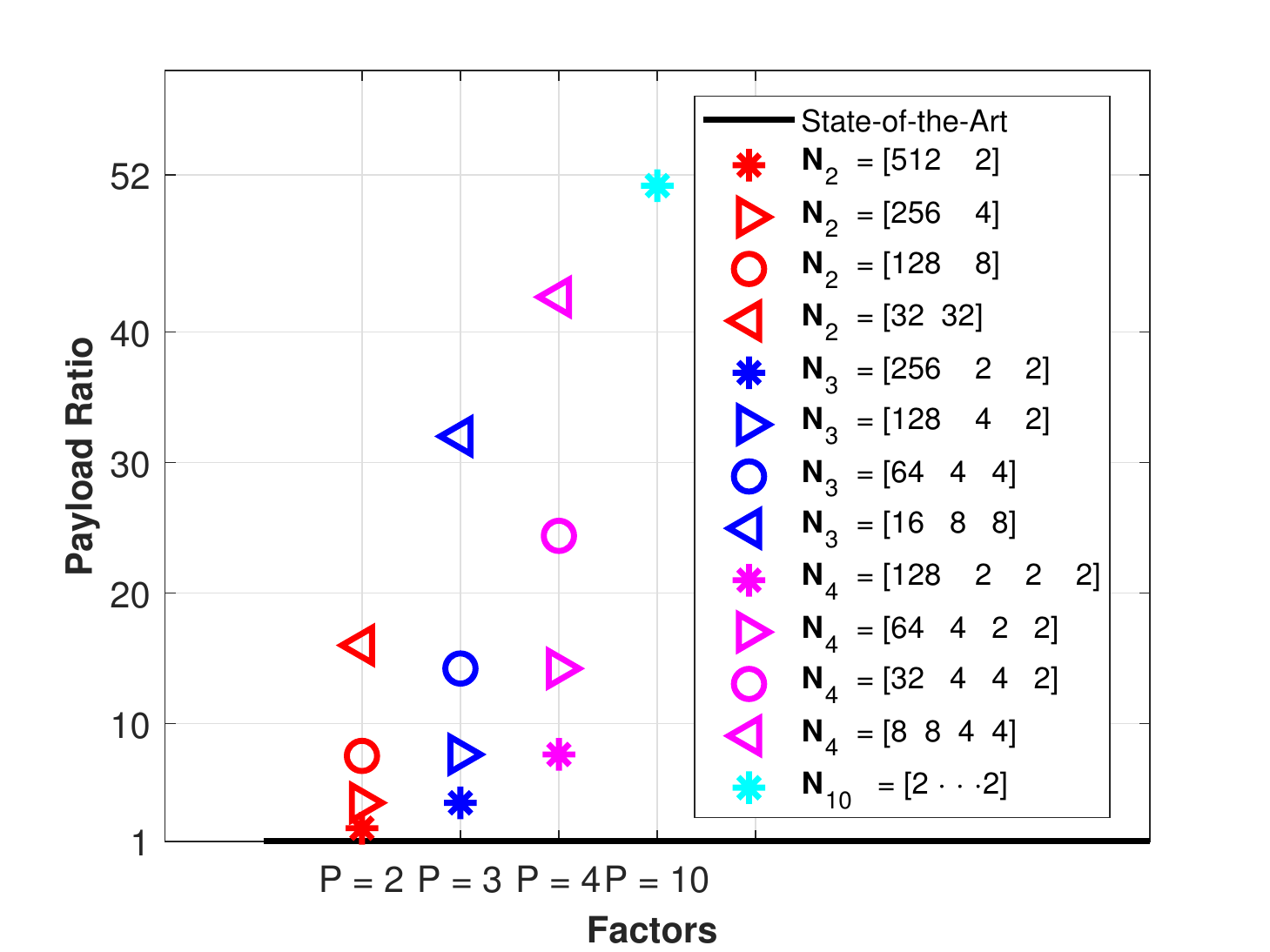}
	\caption{Feedback Payload Ratio for different factors and size configurations. $N=1024$ and $b_\text{F} = b^{(p)}_\text{F} = 3$ bits, for $p = \{1,\ldots,P\}$.}
	\label{fig:payload_ratio}
\end{figure}

Figure \ref{fig:payload_ratio} illustrates the payload ratio (PR) between the proposed feedback overhead model and state-of-the-art, given by  $\text{PR} = N/ \left(\sumx{p=1}{P} N_p \right) $. In this figure, we can observe the role of the $P$ in the factorization model. It allows to decrease the size of the factors, consequentially, allows a massive feedback overhead reduction. For example, in this case where $N=1024$, for $P=2$ we observe a maximum feedback overhead reduction when $\ma{N}_2 = [32,32]$, i.e., we feedback  $64$ phase-shifts against the total of $1024$ phase-shifts that the state-of-the-art does, in other words, a feedback duration $16$ times smaller.  As $P$ increases, we can improve the feedback overhead reduction, at the point that, for $P=10$, the proposed feedback overhead is more than $50$ times smaller than the state-of-the-art \cite{Zappone_Overhead_Aware}.

 \begin{figure}[!t]
	\centering\includegraphics[scale=0.55]{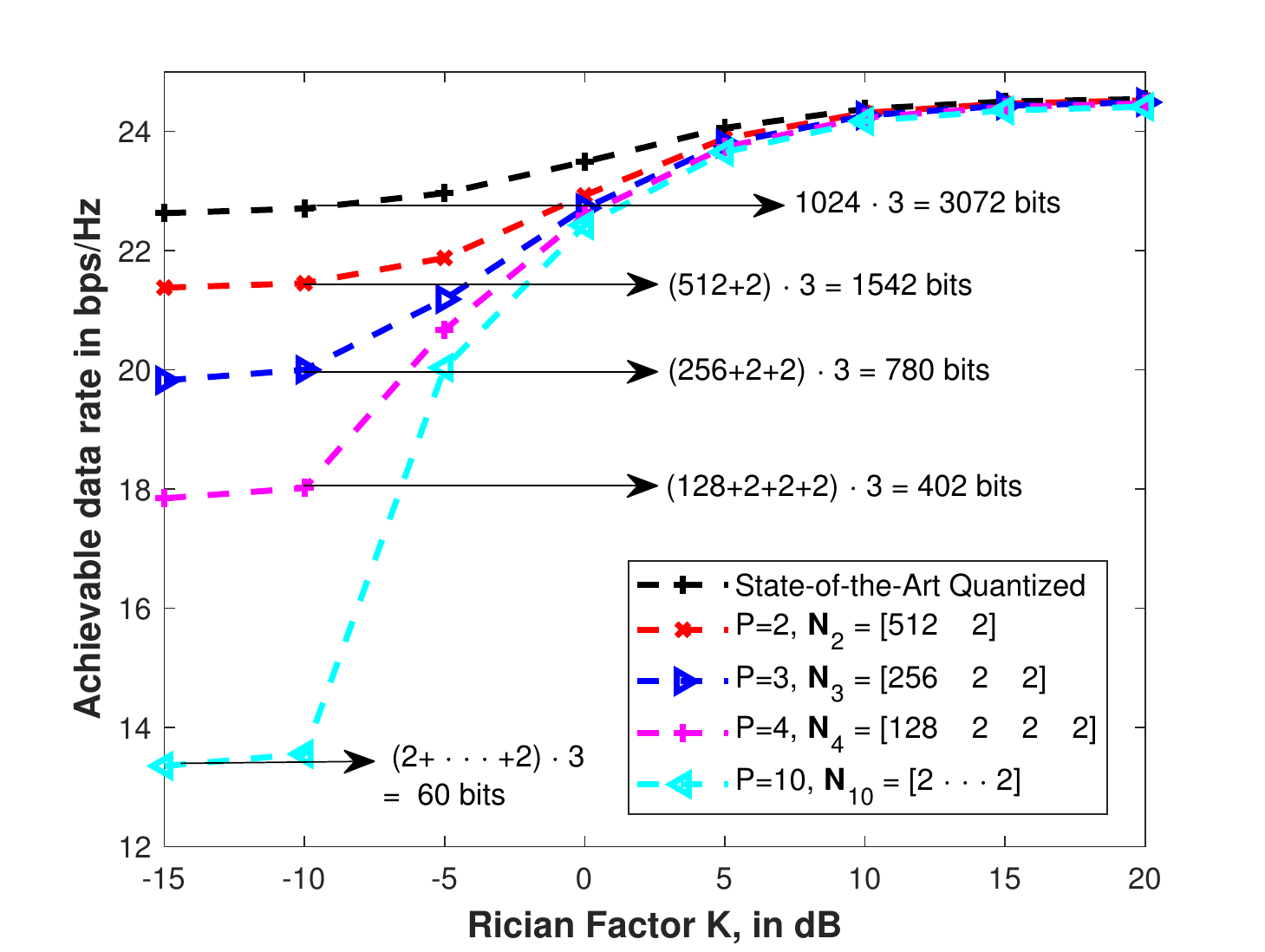}
	\caption{For an IRS with $N=1024$, \ac{TX} and \ac{RX} with $M_R=M_T=2$ and  $ b^{(p)}_\text{F} = b_{\text{F}}= 3$ bits, for the IRS phase-shift quantization resolution,  for $p = \{1, \ldots,P\}$.}
	\label{fig:varying_P}
\end{figure}

 \begin{figure}[!t]
	\centering\includegraphics[scale=0.55]{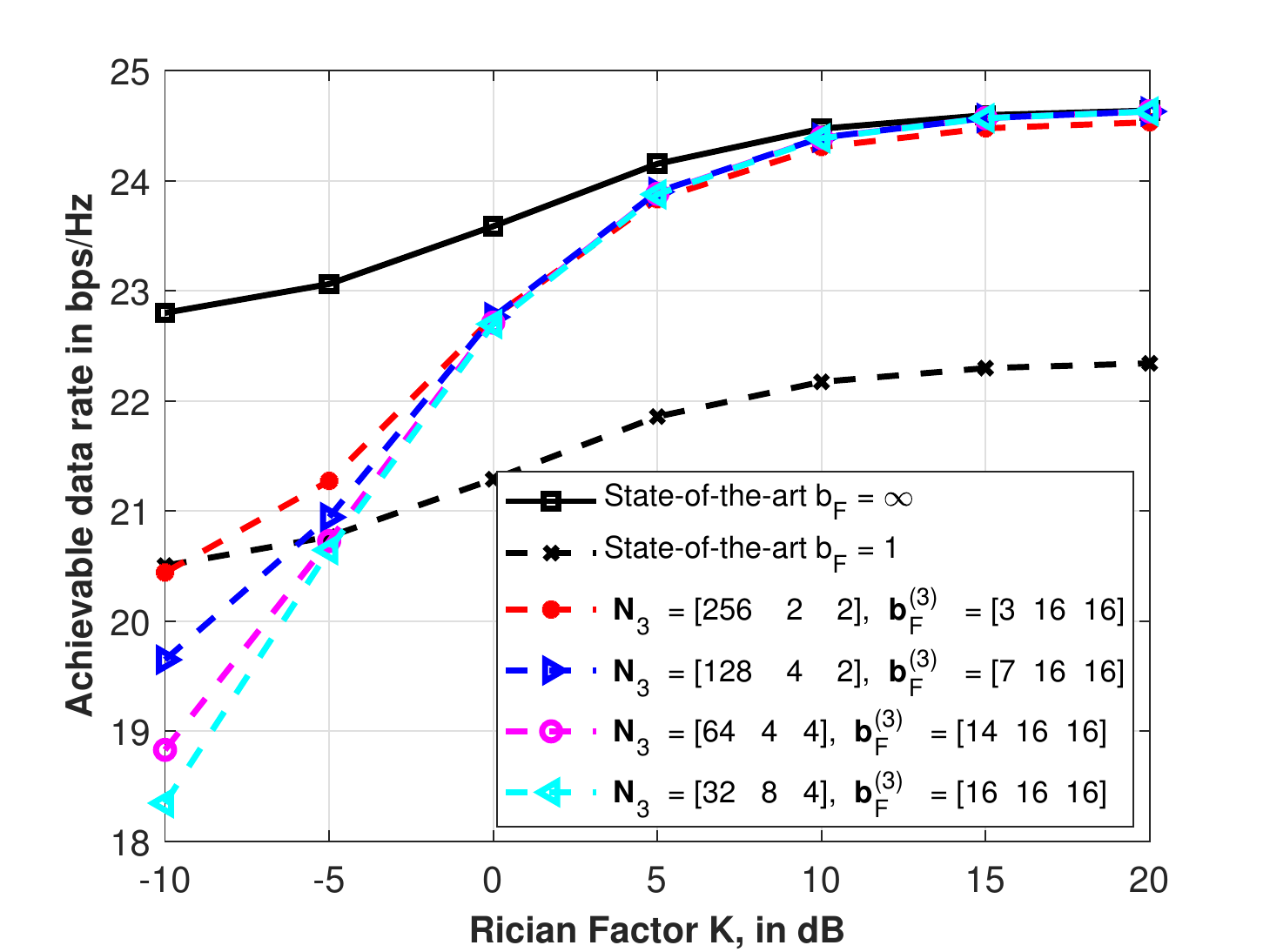}
	\caption{\ac{ADR} of the state-of-the-art and the  proposed factorization for $P=3$ with different size configurations  and resolution (bits), in a fixed control link scenario of $1024$ bits. }
	\label{fig:P_3_fixed_control}
\end{figure} 

In Figure \ref{fig:varying_P}, we study the achievable rate versus the Rician factor $K$ for different factorization methods. Although the proposed IRS phase-shift factorization has a performance degradation in \ac{NLOS} scenarios ($K <0$ dB), the required number of bits for the  feedback of \textcolor{black}{the} IRS phase-shifts is reduced drastically, as also concluded in Figure~ \ref{fig:payload_ratio}.  We also observe that,  as the number of factors $P$ increases the \ac{ADR} decreases. This is explained by the fact that $P$ is the factorization parameter with the role to control the size of each factor, recalling that $N = \prod\limits_{p=1}^P N_p$. In other words, for a higher $P$, the number of independent phase-shifts can be reduced to provide a low feedback payload, while for a smaller $P$ the focus is the performance \ac{ADR} with feedback reduction. \textcolor{black}{Thus, there is a trade-off between the achievable rate and the feedback overhead, based on which one can select the proper factorization parameters.} However, as we move towards \ac{LOS} scenarios ($K> 5$ dB), we can observe that, for all proposed parameter configurations ($P=\{2,3,4,10\}$) we achieve approximately the same data rate while for a higher number of factors ($P=10$) we have a negligible amount of bits to be \textcolor{black}{fed back}, compared to the state-of-the-art. This is explained by the fact that the IRS phase-shift vector $\ma{s} \in \bb{C}^{N \times 1}$ is optimized based on the involved channels $\ma{H}$ and $\ma{G}$ which are geometric-based channels with a separable structure in the IRS (azimuth and elevation), also, \textcolor{black}{every} Vandermond vector can be written as the Kronecker product of multiple factors \cite{sokal2020semi}. Nonetheless, in \ac{LOS} scenarios, a higher $P$ should be prioritized since it has the lowest feedback overhead with a similar \ac{ADR} than the state-of-the-art \cite{Zappone_Overhead_Aware}.

In Figure \ref{fig:P_3_fixed_control} we compare the proposed factorization with the state-of-the-art  for the case of a fixed feedback control link of $1024$ bits. We consider, for the proposed case, $P=3$ factors, with different size configurations and,  consequentially, resolution, i.e., $\ma{N}_{\text{P}}^{\text{T}} \cdot \ma{b}^{(\text{P})}_\text{F} \leq 1024$.  As upper bound, we plot the state-of-the-art with continuous phase-shift (solid curve) to compare with the effect of the quantization. We can observe that, since $N=1024$ phase-shifts, the state-of-the-art can only perform a one-bit quantization, while the proposed method can apply different resolution per factor, in order to meet the control link limit. In this case, for $K \geq -5$ dB, the proposed factorization achieves a better performance than the state-of-the-art, while as $K$ grows, the proposed factorization method achieves a similar performance \textcolor{black}{as in} the state-of-the-art with continuous phase-shift.

To summarize the results illustrated in Figures \ref{fig:payload_ratio}-\ref{fig:P_3_fixed_control}, the  control of the factorization parameters ($P$, $\ma{N}_\text{P}$ and $\ma{b}^{(\text{P})}_\text{F}$) creates a trade-off between \ac{ADR} and feedback reduction (in \ac{NLOS} scenarios), thus \ac{RX} can decide, for example, to feedback the IRS phase-shifts more often, feedback the IRS phase-shifts of multiple users, or in a limited control link channel, can improve the data rate by increase the resolution of the individual factors. \textcolor{black}{ Meanwhile, in the scenarios with a moderate/strong \ac{LOS}, it is preferable factorize the IRS phase-shift vector with the maximum number of factors $P$, which depends of $N$, since $N = \prod\limits_{p=1}^{P} {N_p}$.}

\section{Conclusion and Perspectives}

In this paper, we have proposed a new feedback overhead-aware method by factorizing the optimum IRS phase-shift factor into smaller factors. We have shown the \textcolor{black}{trade-off between the \ac{ADR} and the feedback overhead by varying the factorization parameters}. In a future work, we will  investigate different low-rank approximations models and their trade-off between \ac{ADR} and feedback overhead.

\appendices
\section{LOS Channel Model}\label{App:channel_model}
 The \ac{LOS} components are given as
\begin{align*}
\ma{H}_{\text{LOS}} &= \alpha_H \ma{b}_{\text{IRS}} \cdot \ma{a}^{\text{H}}_\text{TX} \in \bb{C}^{N \times M_T}, \\
\ma{G}_{\text{LOS}} &= \alpha_G \ma{b}_{\text{RX}} \cdot \ma{a}^{\text{H}}_\text{IRS} \in \bb{C}^{M_R \times N}, 
\end{align*}
where $\alpha_H$ and $\alpha_G$ are the path-loss components of the links TX-IRS and IRS-RX, respectively. By considering that the TX and RX are equipped with \acp{ULA}, in which, the spacing between the antennas is $\lambda/2$, their steering vectors, $\ma{a}_\text{TX}$ and $\ma{b}_\text{RX}$ can be written as

\begin{align}
\ma{a}_\text{TX} = \left[1, e^{j\pi \text{sin}\theta_{\text{TX}}}, \ldots, e^{j\pi (M_T -1) \text{sin}\theta_{\text{TX}}} \right]^{\text{T}} \in \bb{C}^{M_T \times 1},  \\
\ma{b}_\text{RX} = \left[1, e^{j\pi \text{sin}\theta_{\text{RX}}}, \ldots, e^{j\pi (M_R -1) \text{sin}\theta_{\text{RX}}} \right]^{\text{T}} \in \bb{C}^{M_R \times 1},
\end{align}
where $\theta_{\text{TX}}$ and $\theta_{\text{RX}}$ are the \ac{TX} and \ac{RX} \ac{AOD} and \ac{AOA}, respectively. Regarding to the  IRS steering vectors on the \ac{AOA} and \ac{AOD} directions,   they are given,  as $\ma{b}_{\text{IRS}} = \ma{b}^{(v)}_{\text{IRS}} \otimes  \ma{b}^{(h)}_{\text{IRS}} \in \bb{C}^{N_hN_v \times 1}$ and $\ma{a}_{\text{IRS}} = \ma{a}^{(v)}_{\text{IRS}} \otimes  \ma{a}^{(h)}_{\text{IRS}} \in \bb{C}^{N_hN_v \times 1}$,  $\ma{b}^{(h)}_{\text{IRS}} \in \bb{C}^{N_h \times 1}$ and $\ma{b}^{(v)}_{\text{IRS}} \in \bb{C}^{N_v \times 1}$  are the \ac{AOA} steering vectors in the azimuth and elevation directions, respectively, and  $N = N_hN_v$. Likewise, $\ma{a}^{(h)}_{\text{IRS}} \in \bb{C}^{N_h \times 1}$ and $\ma{a}^{(v)}_{\text{IRS}} \in \bb{C}^{N_v \times 1}$. The arrival steering vectors are given as 
\begin{align*}
\ma{b}^{(h)}_{\text{IRS}} &= [1, e^{j\pi \text{sin}\psi^{\text{AOA}}_{\text{IRS}}\text{cos}\phi^{\text{AOA}}_{\text{IRS}}}, \ldots, e^{j\pi (N_h -1)\text{sin}\psi^{\text{AOA}}_{\text{IRS}}\text{cos}\phi^{\text{AOA}}_{\text{IRS}}}],  \\
\ma{b}^{(v)}_{\text{IRS}} &= [1, e^{j\pi \text{cos}\phi^{\text{AOA}}_{\text{IRS}}}, \ldots, e^{j\pi (N_h -1)\text{cos}\phi^{\text{AOA}}_{\text{IRS}}}],
\end{align*}
In the same way for the AOD steering vectors $\ma{a}^{(h)}_{\text{IRS}}$ and $\ma{a}^{(v)}_{\text{IRS}}$. All the azimuth \ac{AOA} and \ac{AOD} are generated from a uniform random distribution in $[-\pi, \pi]$, while the elevation \ac{AOA} and \ac{AOD} follows a uniform random distribution within $[0,\pi/2]$.

\begin{thebibliography}{10}
\providecommand{\url}[1]{#1}
\csname url@samestyle\endcsname
\providecommand{\newblock}{\relax}
\providecommand{\bibinfo}[2]{#2}
\providecommand{\BIBentrySTDinterwordspacing}{\spaceskip=0pt\relax}
\providecommand{\BIBentryALTinterwordstretchfactor}{4}
\providecommand{\BIBentryALTinterwordspacing}{\spaceskip=\fontdimen2\font plus
\BIBentryALTinterwordstretchfactor\fontdimen3\font minus
  \fontdimen4\font\relax}
\providecommand{\BIBforeignlanguage}[2]{{%
\expandafter\ifx\csname l@#1\endcsname\relax
\typeout{** WARNING: IEEEtran.bst: No hyphenation pattern has been}%
\typeout{** loaded for the language `#1'. Using the pattern for}%
\typeout{** the default language instead.}%
\else
\language=\csname l@#1\endcsname
\fi
#2}}
\providecommand{\BIBdecl}{\relax}
\BIBdecl

\bibitem{zhang_tutorial}
Q.~Wu, S.~Zhang, B.~Zheng, C.~You, and R.~Zhang, ``Intelligent reflecting
  surface aided wireless communications: A tutorial,'' \emph{IEEE Transactions
  on Communications}, pp. 1--1, May 2021.

\bibitem{jian2022reconfigurable}
M.~Jian, G.~C. Alexandropoulos, E.~Basar, C.~Huang, R.~Liu, Y.~Liu, and
  C.~Yuen, ``Reconfigurable intelligent surfaces for wireless communications:
  Overview of hardware designs, channel models, and estimation techniques,''
  \emph{arXiv preprint arXiv:2203.03176}, 2022.

\bibitem{rajatheva2020scoring}
N.~Rajatheva, I.~Atzeni, S.~Bicais, E.~Bjornson, A.~Bourdoux, S.~Buzzi,
  C.~D'Andrea, J.-B. Dore, S.~Erkucuk, M.~Fuentes \emph{et~al.}, ``{Scoring the
  terabit/s goal: Broadband connectivity in 6G},'' \emph{arXiv preprint
  arXiv:2008.07220}, 2020.

\bibitem{de2022semi}
G.~T. de~Ara{\'u}jo, P.~R.~B. Gomes, A.~L.~F. de~Almeida, G.~Fodor, and
  B.~Makki, ``{Semi-Blind Joint Channel and Symbol Estimation in IRS-Assisted
  Multi-User MIMO Networks},'' \emph{arXiv preprint arXiv:2202.11087}, 2022.

\bibitem{guo2022dynamic}
H.~Guo, B.~Makki, M.~{\AA}str{\"o}m, M.-S. Alouini, and T.~Svensson, ``{Dynamic
  Blockage Pre-Avoidance using Reconfigurable Intelligent Surfaces},''
  \emph{arXiv preprint arXiv:2201.06659}, 2022.

\bibitem{kenneth_wcnps}
K.~B. d.~A. Benício, B.~Sokal, and A.~L.~F. de~Almeida, ``{Channel Estimation
  and Performance Evaluation of Multi-IRS Aided MIMO Communication System},''
  in \emph{2021 Workshop on Communication Networks and Power Systems (WCNPS)},
  2021, pp. 1--6.

\bibitem{Taha2019}
A.~Taha, M.~Alrabeiah, and A.~Alkhateeb, ``Enabling large intelligent surfaces
  with compressive sensing and deep learning,'' 2019, [Online]. Available:
  https://arxiv.org/abs/1904.10136.

\bibitem{khoshafa2021active}
M.~H. Khoshafa, T.~M. Ngatched, M.~H. Ahmed, and A.~R. Ndjiongue, ``Active
  reconfigurable intelligent surfaces-aided wireless communication system,''
  \emph{IEEE Communications Letters}, vol.~25, no.~11, pp. 3699--3703, 2021.

\bibitem{alexandropoulos2020hardware}
G.~C. Alexandropoulos and E.~Vlachos, ``A hardware architecture for
  reconfigurable intelligent surfaces with minimal active elements for explicit
  channel estimation,'' in \emph{Proc. ICASSP}.\hskip 1em plus 0.5em minus
  0.4em\relax Barcelona, Spain: IEEE, 2020, pp. 9175--9179.

\bibitem{emil2019_relay}
E.~Bj{\"o}rnson, {\"O}.~{\"O}zdogan, and E.~G. Larsson, ``Intelligent
  reflecting surface versus decode-and-forward: How large surfaces are needed
  to beat relaying?'' \emph{IEEE Wireless Communications Letters}, vol.~9,
  no.~2, pp. 244--248, 2019.

\bibitem{yang2020intelligent}
Y.~Yang, B.~Zheng, S.~Zhang, and R.~Zhang, ``Intelligent reflecting surface
  meets {OFDM}: Protocol design and rate maximization,'' \emph{IEEE
  Transactions on Communications}, 2020.

\bibitem{Zappone_Overhead_Aware}
A.~Zappone, M.~Di~Renzo, F.~Shams, X.~Qian, and M.~Debbah, ``Overhead-aware
  design of reconfigurable intelligent surfaces in smart radio environments,''
  \emph{IEEE Transactions on Wireless Communications}, vol.~20, no.~1, pp.
  126--141, 2021.

\bibitem{Lathauwer2000bestr1r2}
L.~De~Lathauwer, B.~De~Moor, and J.~Vandewalle, ``On the best rank-1 and
  rank-(r 1, r 2,..., rn) approximation of higher-order tensors,'' \emph{SIAM
  journal on Matrix Analysis and Applications}, vol.~21, no.~4, pp. 1324--1342,
  2000.

\bibitem{golub2013matrix}
G.~H. Golub and C.~F. Van~Loan, ``Matrix computations.''\hskip 1em plus 0.5em
  minus 0.4em\relax JHU press, 2013, ch.~10.

\bibitem{sokal2020semi}
B.~Sokal, A.~L. de~Almeida, and M.~Haardt, ``Semi-blind receivers for {MIMO}
  multi-relaying systems via rank-one tensor approximations,'' \emph{Signal
  Processing}, vol. 166, p. 107254, 2020.

\end{thebibliography}

\end{document}